\documentclass[11pt,a4paper]{article}
\usepackage[margin=1in]{geometry}
\usepackage{graphicx}
\usepackage{amsmath,amssymb}
\usepackage{hyperref}
\usepackage{authblk}
\usepackage{caption}
\usepackage{cite}
\usepackage{siunitx} 
\usepackage{xcolor}   
\usepackage{bm}       
\usepackage{pgf-pie}  
\usepackage[utf8]{inputenc}
\usepackage{lmodern}
\usepackage{subcaption}
\usepackage{tikz}
\usepackage{wrapfig}
\usepackage{verbatim}
\usepackage[section]{placeins}
\usepackage{placeins}

\definecolor{cardinalred}{HTML}{C5050C}    
\definecolor{pinkred}{HTML}{EFA5A5}        
\definecolor{lightgray}{gray}{0.85}        
\definecolor{mediumgray}{gray}{0.5}
\definecolor{white}{rgb}{1,1,1}            

\title{\textbf{Improving TauFinder reconstruction at a 10 TeV muon collider with the MAIA detector concept}\\[0.5em]
\large Presented at the 32nd International Symposium on Lepton Photon Interactions at High Energies (LP2025),\\
Madison, Wisconsin, USA, August 25–29, 2025}

\author[1]{Cyrus Kianian}
\author[1]{Moses Glassman}
\author[1]{Abdollah Mohammadi}
\affil[1]{\small Department of Physics, University of Wisconsin-Madison, United States}
\affil[ ]{\small \texttt{kianian@wisc.edu mglassman3@wisc.edu mohammadi4@wisc.edu}}

\date{November 10th 2025}

\begin{document}

\maketitle

\begin{abstract}
This study aims to improve the \texttt{TauFinder} reconstruction algorithm for the MAIA detector concept. Through this work, we seek to increase the reconstruction efficiency and identification of hadronically decaying $\tau$ leptons. Through our work, we introduce a dynamic signal cone, known as a shrinking cone, which adjusts its size based on the $p_{T}$ of the $\tau$ candidate. In addition to the already studied one charged hadron and three charged hadron decay modes we extend \texttt{TauFinder} to include decays that consist of one charged hadron plus up to two $\pi^{0}$s. Furthermore, we have developed a tagger that prevents electrons being misidentified as 1-prong tau candidates. Applying this tagger results in near-perfect electron rejection with negligible decrease in 1-prong $\tau$ reconstruction efficiency.
\end{abstract}

\section{Introduction}
The MAIA detector concept is designed for a \SI{10}{\tera\electronvolt} muon collider environment and is capable of making high resolution measurements, while mitigating beam induced background (BIB). The detector itself is comprised of an all-silicon tracker immersed in a 5T solenoid field. Surrounding the solenoid are high-granularity silicon-tungsten and iron-scintillator calorimeters, which capture high-energy electronic and hadronic showers, respectively, and support particle-flow reconstruction. The outermost subsystem comprises an air-gap muon spectrometer, which enables standalone track reconstruction for high-momentum muons [1]. Such a collider-detector pairing will empower high-energy physicists to probe the unknown, enabling studies from high-precision Higgs measurements to searches for beyond Standard Model signatures.

In order to assess MAIA's potential, benchmarks and calibration steps are taken to understand it further. Ideally, demonstrating that MAIA has the capability to reconstruct an elusive particle, such as a $\tau$, serves as a full test of the detector's reconstruction potential, which is crucial when conducting real-world experiments.  

$\tau$ leptons are of particular interest, being the heaviest lepton it can decay hadronically and leptonically, allowing us to test reconstruction performance on charged tracks and neutral particles simultaneously. 
Moreover, $\tau$ leptons play a central role in probing Higgs couplings and electroweak interactions, making them invaluable to such studies.
The $\tau$ lepton's mean lifetime is \(\approx10^{-13}\)s, which is too short to reach the detector itself. Therefore, instead of trying to tag the $\tau$ lepton, reconstruction begins from the decay products that reach the detector, for hadronically decaying $\tau$s this consists of both charged and neutral hadrons.
\section{Methodology}
\noindent
\begin{minipage}[t]{0.6\textwidth} 
\subsection{$\tau$ Reconstruction Study}
\small
Using the latest container provided by muoncollidersoft, 15,000 single Monte Carlo (MC) $\tau$ events are generated using a particle gun and written to a \texttt{.slcio} file. The generated $\tau$s are distributed uniformly over the parameters listed in Fig.~\ref{fig:tau_sim}(a). 
\par
\setlength\parindent{22pt}After generation, the events go through a GEANT4-based simulation (DDSim) utilizing the most recent MAIA detector geometry, following the branching ratios in Fig.~\ref{fig:tau_sim}(b). Next, the simulated hits are converted to realistic hits in the digitization step, which is based on the Key4hep software stack. Now, the realistic hit information is passed to the Pandora Particle Flow Reconstruction Algorithm (PandoraPFA), which utilizes ACTS to convert hits into Particle Flow Objects (PFOs). 
\par
\setlength\parindent{22pt}Finally, Marlin runs the reconstruction processors. The primary processor, \texttt{TauFinder}, uses the PFOs are the basis for reconstruction, acting as the $\tau$ decay products. \texttt{TauFinder} selects the highest energy charged particle as a $\tau$ seed forming and adds any particles inside the signal cone to the $\tau$ candidate, where the cone is defined by the opening angle between the momenta of the two particles, and this process repeats until no further seed is found [2]. Once the algorithm finishes it applies quality cuts on the reconstructed $\tau$ candidates, as shown in Fig.~\ref{fig:tau_sim}(c). Candidates that pass the criteria are added to a collection contained in the final \texttt{.slcio} output file. 
\subsection{Electron Rejection Study}
\small
\label{S2.2:Electron Parameters}
Similar to the $\tau$ study, 10,000 MC electron events are generated over the parameters in Fig.~\ref{fig:tau_sim}(a). The simulation and reconstruction steps are the same as the $\tau$ study, and the default \texttt{TauFinder} algorithm is used. 
\end{minipage} 
\hfill
\begin{minipage}[t]{0.35\textwidth} 
\vspace{40pt}
\raggedright
\noindent\textbf{Figure 1:}
\refstepcounter{figure} 
\label{fig:tau_sim}    
\par
\small(a) Monte Carlo $\tau$ Parameters:
\begin{tabular}{lcc}
\hline
Parameter & Min & Max \\
\hline
$\phi$ [rad] & 0 & $2\pi$ \\
$\theta$ [°] & 10 & 170 \\
$p_{T}$ [GeV/c] & 20 & 320 \\
\hline
\end{tabular}

\vspace{1mm} 

\small(b) Simulated $\tau$ Branching Ratios:
\begin{center}
\begin{tikzpicture}
\pie[
    text=legend,
    radius=1.5,
    after number=\%,
    color={cardinalred, white, pinkred, mediumgray, lightgray, cardinalred!70}
]
{
12.3/$\pi^- \nu_\tau$,
28.4/$\pi^- \pi^0 \nu_\tau$,
9.8/$\pi^- \pi^0 \pi^0 \nu_\tau$,
10.9/$\pi^- \pi^+ \pi^- \nu_\tau$,
19.4/$\nu_e\, e \,\nu_\tau$,
19.2/$\nu_\mu \,\mu \,\nu_\tau$
}
\end{tikzpicture}
\end{center}

\vspace{1mm} 

{\small{(c) Default \texttt{TauFinder} Criteria:}}
\begin{tabular}{lcc}
\hline
Cut & Min & Max \\
\hline
$M_{inv}$ [GeV/$c^2$] & 0 & 2 \\
$E_{iso}$ [GeV] & 0 & 5 \\
Charged Tracks & 1 & 4 \\
Particles & 1 & 10 \\
\hline
\end{tabular}
\end{minipage} 

\section{Results $\And$ Discussion}
\subsection{Initial Changes $\And$ Shrinking Cone}
In general, $\pi^\pm$ and $\tau$ efficiencies or rates are defined as:
\begin{subequations} \label{eq:efficiencies}
\begin{align}
\begin{aligned}
{\color{black} \boldsymbol{
\epsilon_{\tau} =
\frac{N_{\text{reco, matched}}(\text{mode}, \tau)}
{N_{\text{MC}}(\text{mode}, \tau)}
}}
\mspace{10mu} \text{(a)}
\end{aligned}
\quad\quad
\begin{aligned}
{\color{black} \boldsymbol{
\epsilon_{\pi^{\pm}} =
\frac{N_{\text{reco, matched}}(\pi^{\pm})}
{N_{\text{MC}}(\pi^{\pm})}
}}
\mspace{10mu} \text{(b)}
\end{aligned}
\tag{1.1} \label{eq:epsilon}
\end{align}
\end{subequations}
where $N_{\text{reco, matched}}$ is the number of reconstructed and MC-matched particles of the specified type (either $\tau$ or $\pi^\pm$). For the $\tau$s the parameter ``mode'' denotes the type of decay. Then, $N_{\text{MC}}$ represents the total number of MC particles of the specified type, where ``mode'' is the MC $\tau$'s decay. So, for a reconstructed 3-Prong (3P) $\tau$ to be counted, it must match to a MC 3P $\tau$ and have exactly 3 reconstructed $\pi^\pm$. Note the $\pi^\pm$s, in green, act as a theoretical maximum since a $\tau$ cannot be reconstructed properly when missing reconstructed $\pi^\pm$. 
\par
\begin{figure}[!h]
  \centering
  \begin{minipage}[t]{0.45\textwidth}
    \centering
    \includegraphics[width=\textwidth]{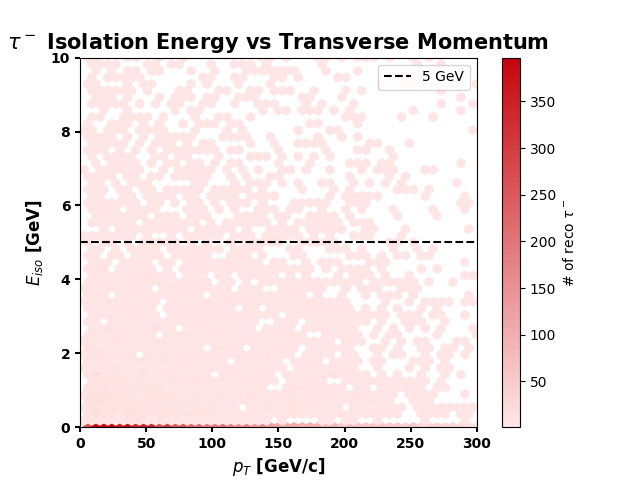}
    \caption{$\tau$ $E_{iso}$ vs. true $p_{T}$.}
    \label{fig2:tau_v_Eiso}
  \end{minipage}
  \begin{minipage}[t]{0.45\textwidth}
    \centering
    \includegraphics[width=\textwidth]{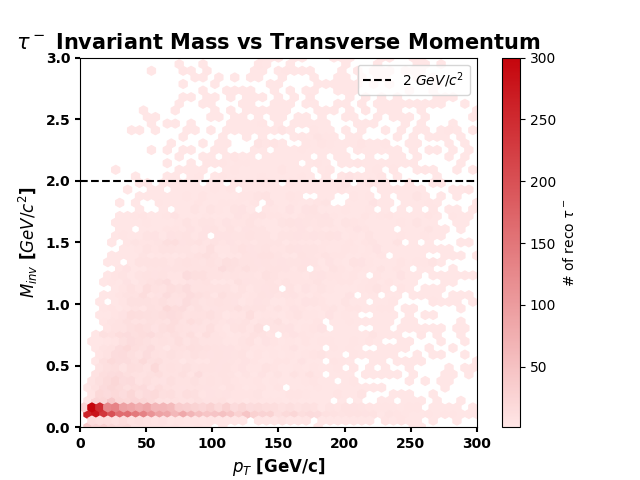}
    \caption{$\tau$ $M_{inv}$ vs. true $p_{T}$.}
    \label{fig3:tau_v_Minv}
  \end{minipage}
\end{figure}

The first studies examining the default \texttt{TauFinder} quality cuts revealed that the $E_{iso}$ cut, set at \SI{5}{\giga\electronvolt}, was responsible for rejecting nearly 14$\%$ of all reconstructed $\tau$s (see Fig.~\ref{fig2:tau_v_Eiso}). Where $E_{iso}$ is the collective energy of the particles within +0.02 rad outside the signal cone. Additionally, the default $M_{inv}$ cut, set at \SI{2}{\giga\electronvolt}/$c^2$, rejected approximately 10$\%$ of all reconstructed $\tau$s (see Fig.~\ref{fig3:tau_v_Minv}). Where $M_{inv}$ is the invariant mass of the reconstructed $\tau$. However, for a non-BIB environment these cuts are irrelevant and hence, set such that they are negligible. In the subsequent plots we've dubbed these changes ``loose'' and colored them blue.
\vspace{1cm}
\par
\begin{figure}[!h]
  \centering
   \begin{minipage}[t]{0.47\textwidth}
    \centering
    \includegraphics[width=\textwidth]{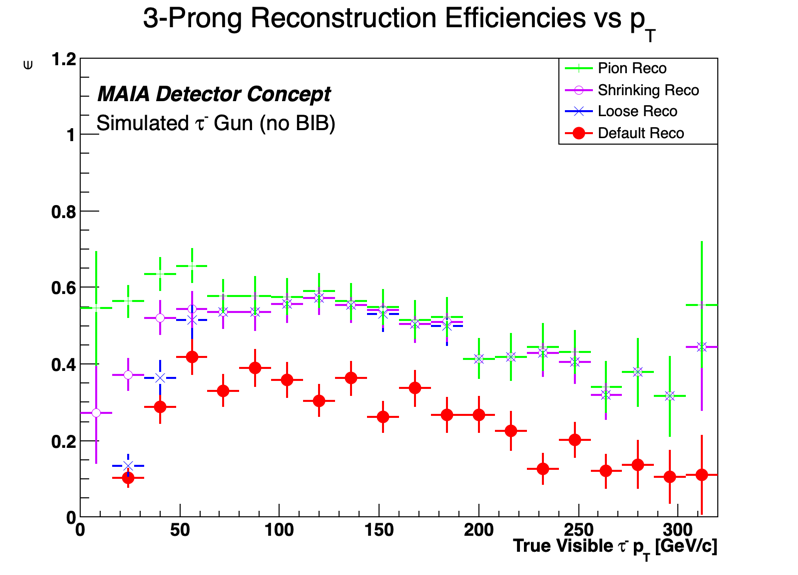}
    \caption{3P reconstruction efficiency vs. $p_{T}$.}
    \label{fig4:3-Prong_eff}
  \end{minipage}
  \begin{minipage}[t]{0.47\textwidth}
    \centering
    \includegraphics[width=\textwidth]{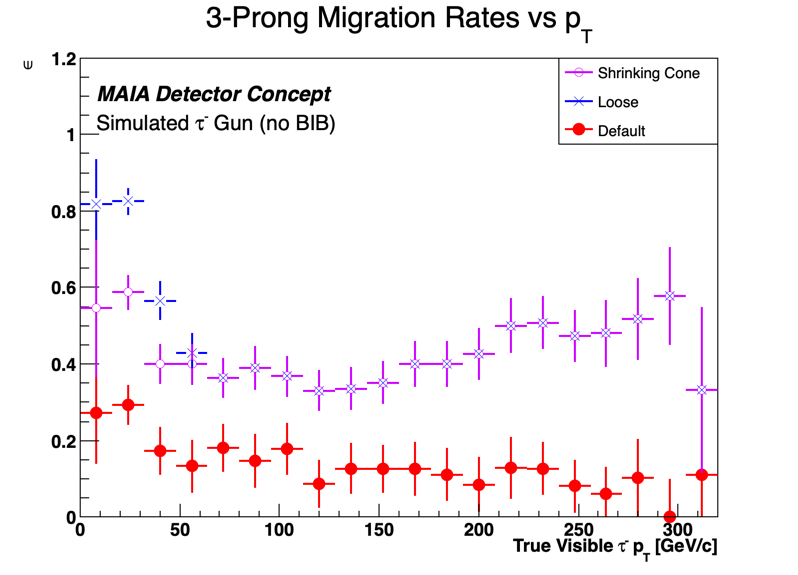}
    \caption{3P migration rate vs. $p_{T}$.}
    \label{fig5:3p_migration}
  \end{minipage}
\end{figure}

The loose cuts recover a considerable amount of efficiency ($\sim15\%$ in the 3Ps, see Fig.~\ref{fig4:3-Prong_eff}), compared to the default \texttt{TauFinder} criteria (red). However, they fail to do so at low $p_{T}$ ($\lesssim$ \SI{60}{\giga\electronvolt}/c). Furthermore, Fig.~\ref{fig5:3p_migration} shows a large increase in low $p_{T}$ 3P migration rate, more so than higher values. Migration rate is defined as Eq.~\ref{eq:epsilon}(a) where the numerator's mode does not match the denominator's mode (3P in this case). A high migration rate indicates a majority of low $p_{T}$ 3P $\tau$s have $\pi^\pm$s laying outside the static search cone size (0.05 rad), motivating a $p_{T}$-dependent shrinking cone. Here it is defined as: 
{
\setlength{\abovedisplayskip}{0pt}  
\setlength{\belowdisplayskip}{2pt}
\[
\text{Signal Cone Angle} =
\begin{cases}
0.3, & p_T \leq 20 \\[0pt]
\dfrac{6}{p_T}, & 20 < p_T < 60 \\[0pt]
0.1, & p_T \geq 60
\end{cases}
\]}where $p_{T}$ is the reconstructed $\tau$'s, which is dynamically updated in \texttt{TauFinder} as PFOs are added to the $\tau$ candidate. In the following results ``shrinking'' refers to the version of the \texttt{TauFinder} algorithm that has both the loose criteria and the shrinking cone modifications; labeled in purple. Together these changes show a reduction of about 20$\%$ in low $p_{T}$ 3P migration rate (Fig.~\ref{fig5:3p_migration}), which corresponds to a similar 3P efficiency increase in the same region (Fig.~\ref{fig4:3-Prong_eff}).

\subsection{New Decay Modes}
Previous work has studied the pure 1-Prong (1P) and 3-Prong cases (those without $\pi^0$), together accounting for 23$\%$ of all simulated decays. However, a majority of simulated decays, about 38$\%$, include $\pi^0$s. So, we encompass the neutral cases to complete the hadronic $\tau$ decays.
\par
By default, PandoraPFA does not reconstruct $\pi^0$, as the decaying $\gamma$s are difficult to resolve. Hence, we utilize a pseudo-$\pi^0$ criteria to study decays with $\pi^0$s. In essence, if a reconstructed $\tau$ has a reconstructed $\gamma$ the $\tau$ is considered to be from a neutral decay (i.e. MC decay modes with a $\pi^0$). Furthermore, we group all neutral MC and reconstructed 1P decay modes into one category, labeled as ``1P+N'', so the number of reconstructed $\gamma$s has does not affect the results.
\par
\begin{figure}[!h]
  \centering
  \begin{minipage}[t]{0.48\textwidth}
    \centering
    \includegraphics[width=\textwidth]{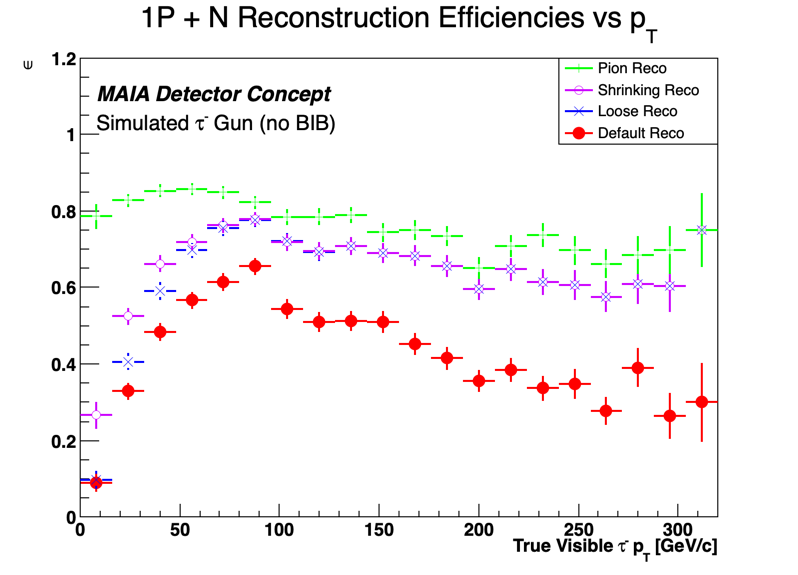}
    \caption{1-Prong + $\pi^{0}$s efficiency vs. $p_{T}$.}
    \label{fig6:1PXN_eff}
  \end{minipage}
  \hfill
  \begin{minipage}[t]{0.5\textwidth}
    \centering
    \includegraphics[width=\textwidth]{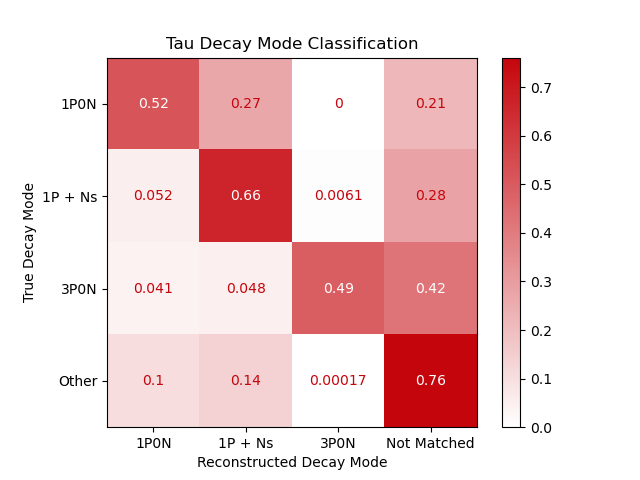}
    \caption{Reconstructed $\tau$ Confusion Matrix.}
    \label{fig7:cm}
  \end{minipage}
\end{figure}

The efficiency of the 1P neutral decay modes is given in Fig.~\ref{fig6:1PXN_eff}, which follows a similar pattern to the 3P efficiency. The loose criteria (blue) greatly improves on the default criteria (red), then combining the shrinking cone with the loose criteria (purple) recovers efficiency at low $p_{T}$, while remaining under the theoretical maximum in the $\pi^\pm$ efficiency (green). 
\par
Fig.~\ref{fig7:cm} shows a confusion matrix for the hadronic decays, with MC or true mode on the left and reconstructed mode on the bottom. The values shown are normalized by the true decay mode, so the number of $\tau$ reconstructed in a given mode are divided by the corresponding number of MC $\tau$ in the same mode. The true decay mode labeled ``Other'' represents the leptonic decays, while the reconstructed ``Not Matched'' mode contains all leptonic decays and hadronic decays that migrated due to a mismatch in the reconstructed and true number of $\pi^\pm$s or $\pi^0$s. From Fig.~\ref{fig7:cm} the 1P+N modes have the best classification rate, indicating the pseudo-$\pi^0$ criteria is a decent approximation (for now). Furthermore, the confusion between true 1P and reconstructed 1P+N modes can be explained by the presence of low-$p_{T}$ $\gamma$s. 
\subsection{Electron Rejection}
Fig.~\ref{fig7:cm} indicates a significant amount of the true, leptonically decaying $\tau$s are reconstructed in the 1P or 1P+N modes. Since electrons and $\mu$s produce one charged track, both are capable of faking $\pi^\pm$s, which \texttt{TauFinder} interprets as a hadronic $\tau$ decay. From the no cut criteria (yellow) in Fig.~\ref{fig12:EEH Electron}, one can see electrons have a $\tau$ fake rate of around 50\%. Overwhelmingly, an electron fakes only a single $\pi$ therefore the following focuses on the faking of 1P or 1P+N $\tau$. 

\begin{figure}
  \begin{minipage}[t]{0.45\textwidth}
    \centering
    \includegraphics[width=\textwidth]{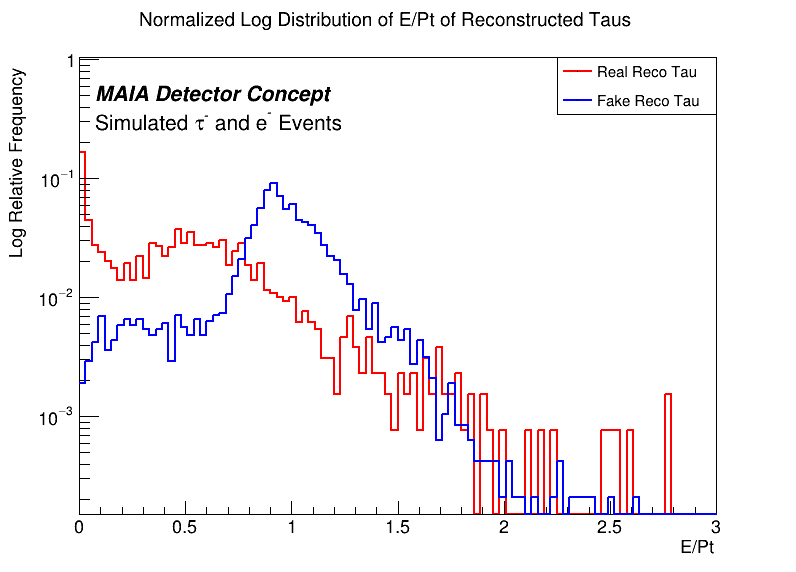}
    \caption{E/Pt for electrons and $\tau$}
    \label{fig10:E/Pt}
  \end{minipage}
  \hfill
  \begin{minipage}[t]{0.45\textwidth}
    \centering
    \includegraphics[width=\textwidth]{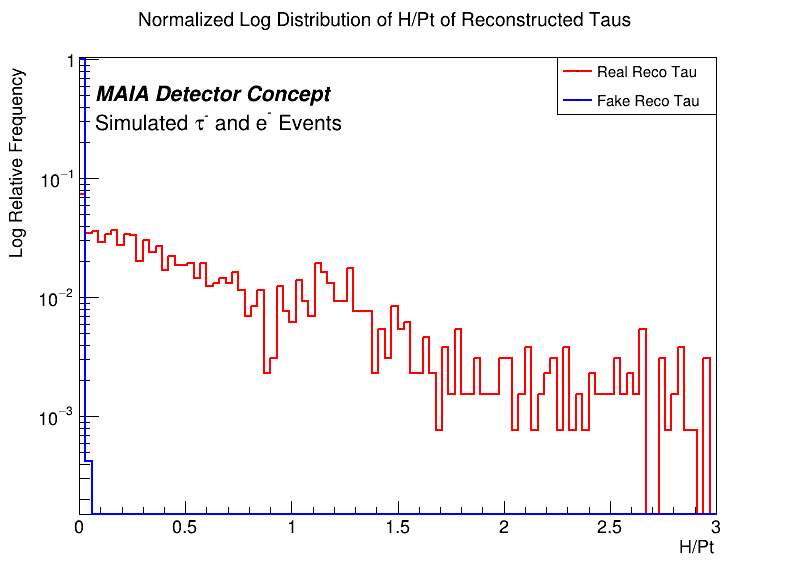}
    \caption{H/Pt for electrons and $\tau$}
    \label{fig11:H/Pt}
  \end{minipage}
\end{figure}
The energy deposition distribution of both $\tau$ and electrons in the electromagnetic calorimeter (ECal/E) and hadron calorimeter (HCal/H) serves as the basis to distinguish hadronically decaying $\tau$ from electrons. As seen in Fig.~\ref{fig10:E/Pt} and~\ref{fig11:H/Pt}, electrons deposit nearly all of their energy in ECal while real $\tau$ split more evenly between ECal and HCal. Constructing a cut based on the ratio \(\frac{E}{E+H}\) of the reconstructed particle, requiring that those below a limit are real $\tau$, yields highly effective results with high benchmarks. With \(\frac{E}{E+H} < 0.99\) the electron fake rate drops to 0.06\% (Fig.~\ref{fig12:EEH Electron}). While the cut does harm real $\tau$ reconstruction efficiency (Fig.~\ref{fig13:EEH Tau}), it is to a minute level and within the error bars of the no cut criteria, thus, negligible in its effect.
\begin{figure}
  \begin{minipage}[t]{0.45\textwidth}
    \centering
    \includegraphics[width=\textwidth]{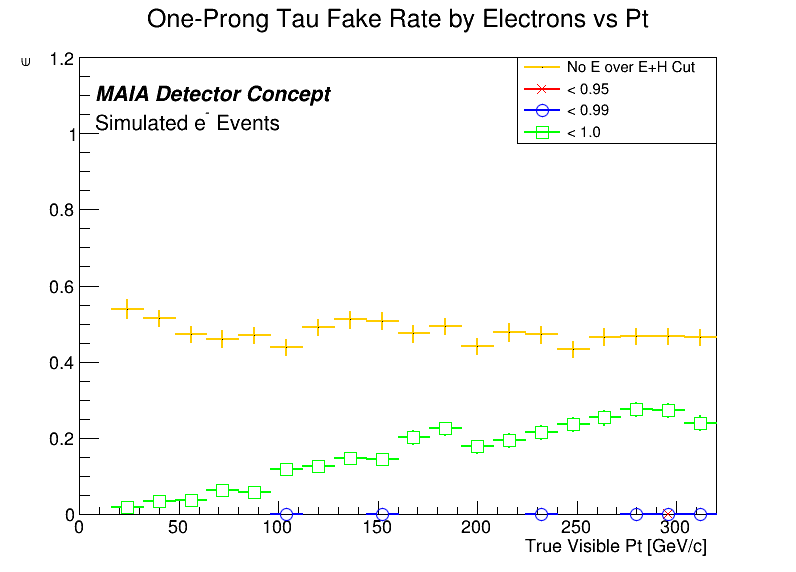}
    \caption{Fake Rate for E/(E+H) Cut}
    \label{fig12:EEH Electron}
  \end{minipage}
  \hfill
  \begin{minipage}[t]{0.45\textwidth}
    \centering
    \includegraphics[width=\textwidth]{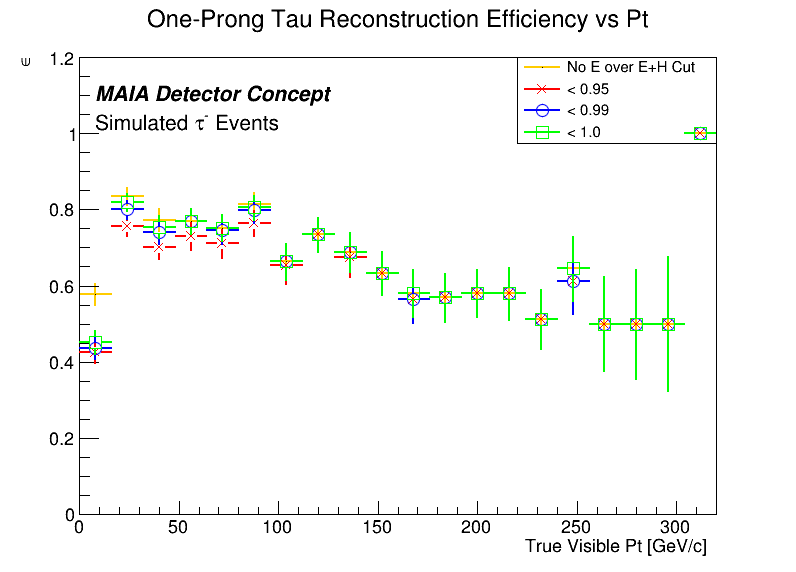}
    \caption{Efficiencies for E/(E+H) Cut}
    \label{fig13:EEH Tau}
  \end{minipage}
\end{figure}

\section{Conclusions}
Through our studies, we demonstrate a $p_{T}$-dependent shrinking cone has the potential to boost low $p_{T}$ ($\lesssim$ \SI{60}{\giga\electronvolt}/c) $\tau$ reconstruction efficiency. In addition, extend \texttt{TauFinder} to reconstruct hadronic decays with $\pi^0$, by using a reconstructed $\gamma$ as a pseudo-$\pi^0$. Furthermore, we achieve near-perfect electron rejection with negligible decrease in the reconstructed 1P $\tau$ efficiency, by applying a cut on a charged particles ECal and HCal energies. Thus, eliminating nearly all of the electron confused with $\pi^\pm$.
\par
Moving forward, we will fine-tune our cuts and shrinking-cone definition with BIB overlaid on the events. Additionally, the pseudo-$\pi^0$ criteria needs further studies. Finally, the Pandora reconstruction algorithm often confuses electron and $\mu$ with $\pi^\pm$, which must be addressed. 

\enlargethispage{\baselineskip}  
\vspace{-1em}
\vspace{-0.5em}
\section*{Acknowledgments}
Thank you to the UW-Madison CMS collaboration for hosting the Lepton-Photon 2025 Symposium in addition to their guidance and feedback on this project. Thank you to our collaborators at Yale and LIP for their commitment and contributions to \texttt{TauFinder}. Thank you to the International Muon Collider Collaboration for their continued support of our research and their relentless pursuit toward the next generation of particle colliders. This work was generously supported by the U.S. DOE.

\vspace{-1.5em}
\bibliographystyle{unsrt}

\end{document}